\documentclass[conference]{IEEEtran}
\ifCLASSINFOpdf
\else
\fi
\usepackage{fixltx2e}
\usepackage{stfloats}
\usepackage{url}
\begin{document}
%
% paper title
% can use linebreaks \\ within to get better formatting as desired
\title{A Study on Optimized Resource Provisioning in Federated Cloud}
% author names and affiliations
% use a multiple column layout for up to three different
% affiliations
\author{\IEEEauthorblockN{Thiruselvan Subramanian\IEEEauthorrefmark{1} and Nickolas Savarimuthu\IEEEauthorrefmark{2}}
\IEEEauthorblockA{Department of Computer Applications,\\
National Institute of Technology,\\
Tiruchirappalli - 620015, INDIA.\\
Email: thirulic@gmail.com\IEEEauthorrefmark{1}, nickolas@nitt.edu\IEEEauthorrefmark{2}} }

% conference papers do not typically use \thanks and this command
% is locked out in conference mode. If really needed, such as for
% the acknowledgment of grants, issue a \IEEEoverridecommandlockouts
% after \documentclass

% for over three affiliations, or if they all won't fit within the width
% of the page, use this alternative format:
% 
%\author{\IEEEauthorblockN{Michael Shell\IEEEauthorrefmark{1},
%Homer Simpson\IEEEauthorrefmark{2},
%James Kirk\IEEEauthorrefmark{3}, 
%Montgomery Scott\IEEEauthorrefmark{3} and
%Eldon Tyrell\IEEEauthorrefmark{4}}
%\IEEEauthorblockA{\IEEEauthorrefmark{1}School of Electrical and Computer Engineering\\
%Georgia Institute of Technology,
%Atlanta, Georgia 30332--0250\\ Email: see http://www.michaelshell.org/contact.html}
%\IEEEauthorblockA{\IEEEauthorrefmark{2}Twentieth Century Fox, Springfield, USA\\
%Email: homer@thesimpsons.com}
%\IEEEauthorblockA{\IEEEauthorrefmark{3}Starfleet Academy, San Francisco, California 96678-2391\\
%Telephone: (800) 555--1212, Fax: (888) 555--1212}
%\IEEEauthorblockA{\IEEEauthorrefmark{4}Tyrell Inc., 123 Replicant Street, Los Angeles, California 90210--4321}}

% use for special paper notices
%\IEEEspecialpapernotice{(Invited Paper)}

% make the title area
\maketitle

\begin{abstract}
%\boldmath
Cloud computing changed the way of computing as utility services offered through public network. Selecting multiple providers for various computational requirements improves performance and minimizes cost of cloud services than choosing a single cloud provider. Federated cloud improves scalability, cost minimization, performance maximization, collaboration with other providers, multi-site deployment for fault tolerance and recovery, reliability and less energy consumption. Both providers and consumers could benefit from federated cloud where providers serve the consumers by satisfying Service Level Agreement, minimizing overall management and infrastructure cost; consumers get best services with less deployment cost and high availability. Efficient provisioning of resources to consumers in federated cloud is a challenging task. In this paper, the benefits of utilizing services from federated cloud, architecture with various coupling levels, different optimized resource provisioning methods and challenges associated with it are discussed and a comparative study is carried out over these aspects.\\
\\
\keywords Federated Cloud, Resource Provisioning, Optimization. 
\end{abstract}

% IEEEtran.cls defaults to using nonbold math in the Abstract.
% This preserves the distinction between vectors and scalars. However,
% if the conference you are submitting to favors bold math in the abstract,
% then you can use LaTeX's standard command \boldmath at the very start
% of the abstract to achieve this. Many IEEE journals/conferences frown on
% math in the abstract anyway.

% no keywords

% For peer review papers, you can put extra information on the cover
% page as needed:
% \ifCLASSOPTIONpeerreview
% \begin{center} \bfseries EDICS Category: 3-BBND \end{center}
% \fi
%
% For peerreview papers, this IEEEtran command inserts a page break and
% creates the second title. It will be ignored for other modes.
\IEEEpeerreviewmaketitle

\section{Introduction}
% no \IEEEPARstart

In recent years cloud computing gained more popularity, as many prefer to host their applications in cloud due to factors such as reduced capital expenditures and operational overhead, increased IT responsiveness and efficiency, greater business flexibility through an on-demand, pay-as-you-go model that scales with business needs, more choice in providers and access to latest services available in the market. There are many cloud providers in the market such as Amazon EC2 \cite {amazon}, Microsoft Azure \cite {azure}, GoGrid \cite {gogrid}, Rack-space \cite {rackspace} offering everything “\textit{as a service}”.
IaaS (Infrastructure as a Service) is one of the cloud service type wherein consumers would take computing infrastructure on lease from cloud providers. Cloud providers host, run and maintain the computational infrastructure, where they provide infrastructure to consumers as Virtual Machines (VMs). They are offered in different flavors as Virtual Instances (VI) based on configuration of resources such as CPU cycles, memory, network bandwidth and storage that are predefined by providers or customized according to consumer. To deploy their applications, Cloud consumers should load Virtual Machine Images (VMI) in VI, where VMI is composition of operating system along with required software packages to run the applications. Cloud consumer is responsible for maintaining the VMs which are loaded with operating system and application software. They are charged based on instance type, usage, data center location and pricing plan opted (Reservation plan or On demand plan)\cite{gogridpricing,ehpricing,ec2pricing,rackspacepricing}.

Selecting best resources from providers and allocating resource for their needs in large cloud market is a challenging task for consumer. A third party, cloud broker eases task of service selection from cloud providers and acts as intermediate between cloud consumer and cloud provider. They play an important role in getting the information from the cloud providers about services offered, price of the service, service availability etc., and selects the best resources according to the consumer requirements. They provide a uniform interface to access, manage, monitor and asses different cloud provider services irrespective of cloud providers technology. 
An evaluation against commercial clouds demonstrates that compared to single cloud deployment, multi-cloud deployment of VMs improves performance and reduces cost\cite {tordsson2012cloud}.

Cloud provider is considered to have infinite resources but there would always be an upper limit based upon restrictions on available hardware, network bandwidth, etc. When there are many infrastructure offerings provided by a Cloud provider, there arises a situation where the Cloud provider suffers resource exhaustion. Single cloud deployment model leads to several challenges where cloud service is unavailable to customer due to outages, natural disasters and attacks leaving customers with loss of access to services and data. These factors conclude that the usage of multiple clouds achieves better service, QoS and reliability.

Resource provisioning in federated cloud is the process of finding optimal placement schemes for VMs and reconfigure them according to changes in the environment. Provisioning in federated cloud has several challenges due to lack of common standards and confusion prevails over which standardization methods has to be followed.
Various resource provisioning challenges in federated clouds and the state of the art in optimized resource provisioning methods are discussed in chapter IV. Analysis over open issues are discussed in chapter V.

% The rest of this paper has been organized as follows: Federated cloud benefits, coupling level of VMs and federated cloud architecture are discussed in Section II. Resource provisioning challenges are presented in Section III. In Section IV, different optimization methods for provisioning of resources in federated cloud are discussed. Section V provides an analysis of resource provisioning methods. Finally, conclusions drawn from this survey are stated in Section VI.

\section{Federated Cloud}

Federated cloud is a composition of two or more clouds that remain as distinct entities but are bound together by standardized or proprietary technology that enables data and application portability \cite {liu2011nist}.
Federated cloud opens the door to a range of useful scenarios that take advantage of cloud capabilities such as
\begin{itemize}
\item Using multiple clouds for different applications to match consumer needs.
\item Utilizing public cloud when internal resources are not enough to meet demands.
\item Moving an application between public clouds or internal data center to meet requirements at different stages in its life cycle.
\item Allocating different elements of an application to different environments, \textit{application stretching}.

\end{itemize}
\subsection{Benefits}

Federated cloud brings out the following benefits over single cloud deployment.

\subsubsection{Scalability}

As the demand scales up, cloud bursting addresses peak demand of consumers when they ran out of computing resources; they burst their workload to external cloud on demand basis and pay per use. Cloud providers collaborate with other cloud providers to share their infrastructure with each other through contracts or framework agreements which defines control over resources, Quality of Service(QoS) rendered and availability, when there is demand for more resources.

\subsubsection{Multi-site Deployment}

Allows infrastructure aggregation across distributed data centers for availability of services; minimize data transfer cost by enabling service in data centers in non-peak hours; meet local jurisdiction and regulation on data placement within certain physical boundary.

\subsubsection{Reliability} 

Replication of service in multi-site deployment enables fault tolerance with high availability of services such as data backup, natural disaster recovery and minimal downtime when a site fails.

\subsubsection{Performance and Cost} 
Deployment of services closer to end users increases the performance by reducing the data traffic and improves response time. Dynamic placement of applications enables reduced overall deployment cost by deploying application in cloud provider, which offers service at minimal cost. Collaboration within cloud providers allows providers to scale up their services without spending high on infrastructure. Minimized energy consumption by server consolidation and placement of VMs in energy efficient data centers which in turn minimizes cost spend on energy.

\subsection{Coupling Levels}
Federated cloud exhibits various level of coupling between cloud instances such as level of cooperation among cloud instances, remote resource control, monitoring, security and access levels, possibility of cross-site networks and migrating VMs among cloud instances. Vozmediano et al. \cite {6165242} discuss on federated cloud and explained more on coupling levels of cloud instances and federated cloud architectures. Coupling level of cloud instances in federated cloud environment are classified into three categories and are differentiated in Table I.
\begin{table*}[!t]
\caption{Coupling Level}
\renewcommand{\arraystretch}{1.3}

\centering
\begin{tabular}{|p{2.5 cm}|p{3.5 cm}|p{3.5 cm}|p{3 cm}|p{3 cm}|}
\hline

\bfseries
Coupling Level &\bfseries Control &\bfseries Monitoring and Accounting &\bfseries Cross-site &\bfseries Security \\ 
\hline \hline 
Loosely Coupled & Basic operations over VMs & Basic virtual resource monitoring & None & Single account representing the organization \\ 
\hline
Partial Coupled & Advanced control over VMs, VMs location and affinity constraints & Advanced virtual resource monitoring & Virtual networks and storage & Framework agreements \\ 
\hline
Tightly Coupled & Placement on specific resource, same VI type & Physical resource consumption & Live migration and high availability & User space sharing \\ 
\hline

\hline
\end{tabular}
\end{table*} 
\subsubsection{Loosely Coupled Federation:} 
In this category, cloud instances have less inter-operation among themselves and is suitable for loosely coupled applications. Monitoring is limited and they can perform only basic operations on VMs such as start, stop and resume. Advance features such as cross-site networks and VMs migration between cloud sites and providers are not supported.
\subsubsection{Partially Coupled Federation}

This consists of different cloud partners involved in contract or framework agreement with terms and conditions describing access methods for remote resources. The framework agreement enables certain level of control over remote resources, detailed monitoring and some advanced networking features among partner clouds.

\subsubsection{Tightly Coupled Federation}

This category is usually formed by clouds belonging to the same organization, and normally governed by the same Cloud OS type. Cloud instance can have advanced control over remote resources, access to all the monitoring information of remote resources and can support features such as creation of cross- site networks, cross-site migration of VMs, implementation of high-availability techniques among remote cloud instances, creation of virtual storage systems across site boundaries.

\subsection{Federated Architectures}

Federated cloud has been implemented using different architectures with various coupling level of resources. Basic architectures of federated cloud\cite {6165242} has been summarized in Table II.

\subsubsection{Cloud Bursting Architecture}

When a cloud consumer ran out of computational resources from their own internal data-center or a private cloud, they bursts workload to cloud providers (Public Clouds) to meet their demand for which they will be charged on pay per use basis. If an organization has a VMware \cite{vmware} based internal data center, CloudSwitch \cite{cloudswitch} allows the consumers to seamlessly move their workloads to public clouds such as Amazon EC2\cite{amazon}, Rackspace\cite{rackspace},etc. CloudSwitch, utilizing their \textit{Cloud Isolation Technology}, make the IP and MAC addresses of the public cloud appear as if they belong to an internal network. This allows the applications to run without any change, when moved to public clouds and the workload can be brought back into the internal data center whenever the consumer wishes.

\begin{table*}[!t]
\renewcommand{\arraystretch}{1.3}
\caption{Federated Architecture}
\centering
\begin{tabular}{|p{2.5 cm}| p{5.5 cm}| p{5.5 cm} |p{2.5cm}|}
\hline
\bfseries Cloud Architecture &\bfseries  Cloud Type &\bfseries  Aim &\bfseries  Coupling level\\

\hline\hline

Cloud Burst & Private cloud to scale out with public or virtual private cloud resources & Meet peak demands & Loosely Coupled\\ 
\hline
Cloud Broker & User of several public clouds & Cost, performance and reliability optimization & Loosely Coupled\\ 
\hline
Aggregated Cloud & Aggregation of different private and public clouds & Sharing of resources between partners to meet peak demands & Partial Coupled\\ 
\hline
Multi-Site Cloud & Very large corporate clouds (private, public or virtual private) with several instances & Scalability, isolation or multiple-site support & Tightly Coupled\\ 

\hline
\end{tabular}
\end{table*}

\subsubsection{Cloud Broker Architecture}

Integration and selection of cloud services from different cloud providers is too complex for cloud consumers to manage. Cloud consumers contact cloud broker for cloud services, instead of contacting cloud providers directly. Broker act as pairing service between cloud consumers and cloud providers. They negotiate contracts with cloud providers on behalf of the customer and distribute workloads across multiple cloud providers in an effort for cost effective deployment by hiding the complexity of negotiations. Customers are provided with an application program interface and unified management interface for utilizing cloud services from multiple cloud providers. Cloud broker architecture is loosely coupled with no advanced control over virtual resources of cloud providers. 

\subsubsection{Aggregated Cloud Architecture}

Cloud is believed to have infinite resources but there would always be an upper limit based upon restrictions on available hardware, network bandwidth, etc. When there are many infrastructure offerings provided by a Cloud provider, there arises a situation where cloud provider end up in resource exhaustion. Under such situation a single cloud provider by itself, may not be able to fulfill additional infrastructure requirements of user in peak demands. Aggregated cloud architecture contributes beneath the layers to fulfill the infrastructure requirements of customers seamlessly. Cloud providers inter-operate and aggregate their resources based on contracts or framework agreements to provide their users with a larger virtual infrastructure. The customers would be aware of only the primary cloud provider. Through the primary cloud provider's single virtual interface, the consumers must be able to add virtual infrastructure components and administer them as per their requirements. The actual action (addition, migration, etc.,) on infrastructure components in different clouds would be handled by cloud providers automatically. Cloud providers have advance control over cloud instance they utilize to meet demands as specified in the framework agreements between cloud providers\cite {6165242,infosys}. This architecture is partially coupled and RESERVOIR\cite{rochwerger2009reservoir} is an example of aggregated cloud architecture. 
\subsubsection{Multi-site Cloud Architecture}
Multi-site cloud architecture is implemented in organizations having geographically distributed cloud infrastructures or data center. They have full control over infrastructures and exposed as a single cloud to consumers. The cloud instances are tightly coupled and can perform cross-site networking and live migration of VMs.
\section{Resource Provisioning}

Resource provisioning in IaaS cloud is the process of providing computational infrastructure (i.e. VI) to cloud consumers for their computational needs. Static resource provisioning based on peak demand is not cost-effective because of poor resource utilization during off-peak periods\cite{hu2009resource}. Automatic resource provisioning would lead to efficient resource utilization since, additional resources are provisioned when demand increases and de-provisioned when demand is decreases. 
\subsection{Resource Provisioning Challenges in Federated Cloud}

Integration of various clouds makes resource provisioning a challenging task.
\subsubsection{Application Architecture}

Architecture of an application or a service has to be considered while provisioning of resources and constructing deployment plan in federated environment. Loosely coupled applications can be deployed in multiple clouds; constrains over selection of resources and deployment is minimal. Application components that are less independent, having more coordination and more information flow with other components (i.e. tightly coupled applications) should be deployed in single cloud or same physical server. 
\subsubsection{Portability and Interoperability}

In federated cloud architecture there should be ability to 
\begin{itemize}
\item Move data, applications, and VMs from one cloud computing environment to another.
\item Mix and match cloud services depending on business need.
\item Blend public and private cloud environments into hybrid cloud.
\item Develop and manage cloud services via industry standard APIs.
\end{itemize} Portability is just not about avoiding vendor lock-in but considering factors such as performance, availability, QoS, scalability, budget and business agility.
There are orchestration tools that can provide higher levels of abstraction for automating portability between clouds without relying on a common cloud API, but there are common issues on portability and interoperability among cloud providers.
They are
\begin{itemize}
\item Integrating different services from one or more cloud service providers.
\item Managing security and business continuity risk across several cloud providers.
\item Managing life cycle of a service in a distributed multiple-provider environment in order to satisfy Service Level Agreement (SLA).
\item Maintaining effective governance and audit processes across integrated data centers and cloud providers.
\item Adopting or switching to a new cloud provider without QoS degradation.
\end{itemize}
Industry standards for portability and interoperability such as the Open Virtualization Format [DMT09] \cite{crosby2010open} and Cloud Data Management Interface [SNI10] \cite{cdmi} are accepted and followed by many industry leaders in cloud.

\subsubsection{Deployment Plan}

In federated cloud, cloud providers offer different services with different pricing models and service quality. Selection of best resources based on application requirements to minimize the budget of customer and maximize the resource performance such as utilization, availability, reliability and minimum response time is a difficult task. Different models on optimized provisioning on resources based on cost and performance are discussed in \textit{Section 4}. Due to uncertainty over demand from customer, availability of resources and price among providers, optimized deployment plan under uncertainty is challenging task \cite{buyya2010intercloud}. Locality of data center determines speed of delivery, streaming and dynamic content to end users, and hence it is better to have data centers located near the end user.
\subsubsection{Quality of Service}

QoS is one of the main features of cloud computing. To maintain the QoS, SLA plays an important role where, SLA is a part of service contract between cloud providers and consumers that define level of service in a formal way. In federated cloud computing environment, QoS is maintained during peak demands by provisioning resources from other clouds if resource is exhausted, recovery from failures by replicating consumer application in different clouds or data center across globe or transparently migrating services to other providers with less or no downtime.
\subsubsection{User Specific Constraints}

The user can specify different constraints to deploy their applications that can restrict the deployment decisions. For example, hardware and platform constraints such as type of hypervisor, operating system, etc., affinity constraints such as two or more VMs that need to be deployed in the same physical server or in the same physical cluster, location constraints such as geographical restrictions on data center of provider where applications and data are to be deployed, or SLA constraints such as guaranteed CPU capacity, high operational reliability, etc\cite {6165242}.
\subsubsection{Jurisdiction and Regulation}

Cloud providers have to comply with international, federal or state directives such that data should reside within certain physical boundaries. So, customer can deploy their applications in different cloud providers who comply with regional laws to provide services for particular region\cite{badger2011draft}.
\subsubsection{Resource Pricing and Instance Type}

Pricing scheme of virtual resources varies among cloud providers. Amazon EC2 offers three types of pricing plans\cite{ec2pricing} namely Reserved, On Demand and Spot Instances for provisioning VI.
ElasticHosts\cite{ehpricing} allows users to customize instances and pricing based on instance resources such as CPU, memory, disk, SSD and data transfer rate. In ElasticHosts and GoGrid\cite{gogridpricing}, billing is based on hourly, monthly and yearly subscription. Rackspace\cite{rackspacepricing} offers servers based on size of RAM and charge per hour usage. Most of cloud providers charge based on storage, memory and network bandwidth where long term subscriptions yields significant amount of saving to customers, but decision making on subscription is a challenging task. 
\begin{table*}[!t]
\renewcommand{\arraystretch}{1.3}
\caption{Optimized Provisioning of Resources in Federated Cloud}
\centering
\begin{tabular}{|p{2.25 cm}|p{1.25 cm}|p{2.5 cm}|p{5 cm}|p{5 cm}|}
% \hline \hline \multicolumn{5}{|c|}{Coupling Level}\\
\hline 
\bfseries Author &\bfseries  Architecture &\bfseries  Optimization Criteria &\bfseries  Decision Values &\bfseries  Methodology \\
\hline \hline 
Bossche et al. \cite{van2010cost} & Burst & Cost & Demand & Binary Integer Programming \\
\hline
Javadi et al. \cite{javadi2012failure} & Burst & Performance and Cost & Deadline , Performance, Cost and Failure of VMs & Knowledge-Free Approach \\
\hline
Chaisiri et al. \cite{chaisiri2012optimization} & Broker & Cost & Dynamic demand and price of resource & Stochastic Programming \\
\hline 
Tordsson et al. \cite{tordsson2012cloud} & Broker & Cost and Performance & Static demand and Price of resource & Binary Integer Programming \\
\hline
Lucas-Simarro et al. \cite{lucas2012scheduling} & Broker & Cost and Performance & Budget and Performance & Binary Integer Programming \\
\hline
Breitgand et al. \cite{breitgand2011policy} & Aggregated & Profit, Performance and Energy Consumption & QoS, Demand and Framework Agreements & Integer Programming and Greedy LP Rounding heuristic 
\\
\hline

Vecchiola et al. \cite{vecchiola2012deadline} & Aggregated & Performance & Deadline, Performance and QoS & Deadline-Driven Provisioning and Resource Pooling\\
\hline
Wright et al. \cite{wright2012constraints} & Aggregated & Cost and Performance & User constraints such as Location, Budget and application resources & Two-phase Constraints-based Discovery Approach \\
\hline
Calheiros et al. \cite{calheiros2012coordinator} & Aggregated & Profit and Performance & Performance, Reliability and Scalability & Cloud Coordinator Architecture \\
\hline
\hline
\end{tabular}
\end{table*}
\section{Optimized Provisioning of Resources in Federated Cloud}

Cloud providers faces the problem of finding an optimal solution for some criteria such as efficient utilization of existing resources by protecting QoS requirements of users and minimizing the overall budget of data centers. Cloud consumers aims at minimizing the cost of getting best service form providers.

\subsection{Cloud Bursting Architecture}

Bossche et al. \cite{van2010cost} proposes a method to minimize cost of external provisioning in which workloads are outsourced from an internal cloud to public cloud i.e. cloud bursting. Their work mainly deals with deadline-constraint and non migrative workloads, where memory, CPU, and networking are integrated in binary integer programming problem formation. They also provide some experimental insight into scalability and tractability of their formulations. 
Javadi et al. \cite{javadi2012failure} consider the problem of QoS-based resource provisioning in a hybrid cloud computing system where the private cloud is failure-prone and to overcome, hybrid cloud architecture is developed. They proposed brokering strategies in the hybrid cloud system where an organization that operates its private cloud aims to improve the QoS for the user's requests by utilizing the public cloud resources and uses Knowledge-Free Approach. 

\subsection{Cloud Broker Architecture}

Chaisiri et al. \cite{chaisiri2012optimization} discuss on optimization of resource provisioning cost in federated cloud with future demand and price uncertainty. Authors consider reservation and on-demand pricing plan where focus is on minimizing on-demand and over subscription cost. Using Deterministic Equivalent Formulation, Sample-Average Approximation, and Benders Decomposition methods for fast decision making.

Tordsson et al. \cite{tordsson2012cloud} discuss on the cloud brokering mechanism that performs two operations (i) the optimal placement of the virtual resources of a virtual infrastructure across a set of cloud providers, and (ii) management and monitoring of these virtual resources by providing unified management user interface. By considering the demand and price of resources as static, they formed a 0-1 Integer Programming to minimize the cost and maximize the performance. Their experimental results confirm that multi-cloud deployment provides better performance and lower costs compared to the usage of a single cloud. 

Lucas-Simarro et al. \cite{lucas2012scheduling} present a cloud brokering architecture that can work with different scheduling strategies for optimal deployment of virtual services across multiple clouds, based on different optimization criteria (e.g. cost optimization or performance optimization), different user constraints e.g. budget, performance, VI types, placement, reallocation or load balancing constraints and different environmental conditions i.e., static vs. dynamic conditions, VI prices, VI types, service workload, etc. Binary Integer programming formulation is used.
\subsection{Aggregated Cloud Architecture}

Breitgand et al. \cite{breitgand2011policy} addresses the management challenge of efficient provisioning of elastic cloud services with a federated approach. Their placement algorithm aims at maximizing providers profit while protecting QoS delivered to consumers. Integer Liner Programming formulations is used to optimize power saving and load balancing internally in a cloud, as well as to minimize the cost for outsourcing workloads to external partners and demonstrates the integration of placement algorithms with the RESERVOIR \cite{rochwerger2009reservoir}in federated cloud computing.

Vecchiola et al. \cite{vecchiola2012deadline} implemented two algorithms in Aneka platform\cite{vecchiola2009aneka} for (i)Deadline-driven provisioning and (ii) Resource pool selection for deadline- driven provisioning of resource from multiple computing sources such as private cloud, public cloud, clusters, grids and desktop grids ,which is responsible for supporting QoS aware execution of scientific applications and efficiently allocate resources from different sources in order to reduce application execution times.

Wright et al. \cite{wright2012constraints} introduces a two-phase constraints-based approach to a multiple cloud provider environment for discovering the most appropriate infrastructure resources for a given application. In first phase, suitable resources are identified for the application and in second phase, heuristic approach based on cost and/or performance is used for selecting best resources from the identified resources.

Calheiros et al. \cite{calheiros2012coordinator} presents architecture of Cloud Coordinator element from InterCloud \cite{buyya2010intercloud} architecture which represents data centers and brokers in the InterCloud marketplace, and it is responsible for publishing offers and requests for resources, discovering potential providers of resources, and negotiating resources when it is necessary. 

\section{Discussions}

Optimized resource provisioning methods are identified and are summarized in Table III.

In cloud bursting, existing methods mainly focus on performance and cost for selecting public clouds while internal data center is exhausted. Customers should consider the heterogeneous nature of public clouds, where provider specific adapter is needed to access resources from cloud by the internal data center. Existing cloud brokering approaches are not flexible enough to support all kinds of applications and QoS requirements. Application specific brokering has to be developed to identify application specific requirements and thus enhance brokering policies accordingly.

Resource provisioning in loosely and partially coupled architectures have been widely studied and implemented focusing mainly on constraints such as cost, performance and QoS. Resources provisioning polices have to be developed based on application architecture, user specific constraints, reliable deployment plan for fault tolerance and recovering from outages such as network failure, natural disaster, etc.

\section{Conclusion}
In this paper, a study on optimized resource provisioning in federated cloud is made where, the basic architectures of federated cloud and the challenges associated with provisioning of resources are discussed. Finally, existing solutions on optimized provisioning of resources in federated cloud, by evaluating their deployment architectures are analyzed to give better perception. \\
\newline
Issues identified for future work are summarized as follows.
\subsubsection{Customized SLA based on application requirements} Existing SLA approaches are not based on application requirements. Providers have set of SLAs, which mainly deals with the performance and availability of the service and not negotiable for application requirements. Application specific SLA has to be developed to meet customer requirements.
\subsubsection{Optimized Deployment Plan based on user constraints} Existing brokering mechanisms mainly deals with the optimized selection of resources from multiple clouds based on cost and performance. Brokering mechanisms should deal with selection of resources based on application and user requirements such as location of data-center, reliability and QoS to meet uncertainty in demand, performance and cost of resources.
\subsubsection{Uniformity in pricing schemes and configuration of VI} Due to lack of standards in pricing and configuration of VI among cloud providers. It is difficult to choose optimized VI for application requirements and VM migration becomes tedious. Industry standards has to be developed to improve interoperability.\\

A flexible system satisfying both cloud consumer and provider could be achieved by fulfilling the above issues in future work.

% trigger a \newpage just before the given reference
% number - used to balance the columns on the last page
% adjust value as needed - may need to be readjusted if
% the document is modified later
%\IEEEtriggeratref{8}
% The "triggered" command can be changed if desired:
%\IEEEtriggercmd{\enlargethispage{-5in}}

% references section

% can use a bibliography generated by BibTeX as a .bbl file
% BibTeX documentation can be easily obtained at:
% http://www.ctan.org/tex-archive/biblio/bibtex/contrib/doc/
% The IEEEtran BibTeX style support page is at:
% http://www.michaelshell.org/tex/ieeetran/bibtex/
%\bibliographystyle{IEEEtran}
% argument is your BibTeX string definitions and bibliography database(s)
%\bibliography{IEEEabrv}

\begin{thebibliography}{4}



 \bibitem{amazon}
 Amazon Elastic Compute Cloud, 2013, Available: http://aws.amazon.com/ec2/

 \bibitem{azure}
 Windows Azure: Microsoft's Cloud Platform $|$ Cloud Hosting $|$ CloudServices, 2013, Available: http://www.windowsazure.com/

 \bibitem{gogrid}
 Welcome to GoGrid, 2013, Available: http://www.gogrid.com/

 \bibitem{rackspace}
 Open Cloud Computing, Managed Hosting, Dedicated Server Hosting by Rackspace, 2013, Available: http://www.rackspace.com/

 \bibitem{gogridpricing}
 Pricing- GoGrid, 2013,  Available: http://www.gogrid.com/products/pricing

 \bibitem{ehpricing}
 Cloud Hosting Pricing - Cloud Server Cost - Hourly On-Demand Price, 2013, Available: http://www.elastichosts.com/cloud-hosting/pricing 

 \bibitem{ec2pricing}
 Amazon EC2 Pricing, 2013, Available: http://aws.amazon.com/ec2/pricing/

 \bibitem{rackspacepricing}
 Cloud Servers Pricing by Rackspace Cloud Computing and Hosting, 2013, Available: http://www.rackspace.com/cloud/public/servers/pricing/ 

 \bibitem{tordsson2012cloud}
 J. Tordsson, R. Montero, R. Moreno-Vozmediano, and I. Llorente, "Cloud brokering mechanisms for optimized placement of virtual machines across multiple providers," Future Generation Computer Systems, Vol. 28, no.2, pp. 358--367, 2012.

 \bibitem{liu2011nist}
  F. Liu, J. Tong, J. Mao, R. Bohn, J. Messina, L. Badger and D. Leaf, "NIST Cloud Computing Reference Architecture," NIST Special Publication, 2011, 500:292.

 \bibitem{6165242}
R. Moreno-Vozmediano, R. S. Montero, and I. M. Llorente, "IaaS Cloud Architecture: From Virtualized Datacenters to Federated Cloud Infrastructures," Computer, vol. 45, no. 12, pp. 65-72, Dec., 2012.


 \bibitem{vmware}
 VMware Virtualization Software for Desktops, Servers and Virtual Machines for Public and Private Cloud Solutions, 2013, Available: http://www.vmware.com/

 \bibitem{cloudswitch}
 Enterprise Cloud Computing - CloudSwitch, 2013, Available: http://www.cloudswitch.com/

 \bibitem{infosys}
 N. Mehrotra, and  N. Dangwal, (201, Aug.) "Interoperate in Cloud with Federation," [Online]. Available: http://www.infosys.com/engineering-services/white-papers/Documents/interoperate-cloud-federation.pdf

 \bibitem{rochwerger2009reservoir}
B. Rochwerger, D. Breitgand, E. Levy, A. Galis, K. Nagin, I.M. Llorente, R. Montero, Y. Wolfsthal, E. Elmroth, J. Caceres, M. Ben-Yehuda, W. Emmerich, and F. Galan, "The Reservoir model and architecture for open federated cloud computing," IBM Journal of Research and Development , vol.53, no.4, pp.4:1,4:11, July 2009

 \bibitem{hu2009resource}
 Y. Hu, J. Wong, G. Iszlai, and M. Litoiu, "Resource provisioning for cloud computing," In Proceedings of the 2009 Conference of the Center for Advanced Studies on Collaborative Research, ACM, pp. 01--111, 2009.

 \bibitem{crosby2010open}
S. Crosby, R. Doyle, M. Gering, M. Gionfriddo, S. Grarup, S. Hand, M. Hapner, and D. Hiltgen, "Open virtualization format specification," vol. DSP0243 1, no. 0, 2010.

 \bibitem{cdmi}
 Cloud Data Management Interface (CDMI), Version 1.0.2, June, 2012.

 \bibitem{buyya2010intercloud}
 R Buyya, R Ranjan, and R. Calheiros, "Intercloud: Utility-oriented federation of cloud computing environments for scaling of application services," Algorithms and architectures for parallel processing, pp.13--31, 2010.

 \bibitem{badger2011draft}
 L. Badger, T. Grance, R. Patt-Corner, and J. Voas, "Draft cloud computing synopsis and recommendations," NIST Special Publication, 800:146, 2011

 \bibitem{van2010cost}
 R. Van den Bossche, K. Vanmechelen, and J. Broeckhove, "Cost-Optimal Scheduling in Hybrid IaaS Clouds for Deadline Constrained Workloads," Cloud Computing (CLOUD), 2010 IEEE 3rd International Conference on ,  pp.228,235, 5-10, July 2010.

 \bibitem{javadi2012failure}
 B. Javadi, J. Abawajy, and R. Buyya, "Failure-aware resource provisioning for hybrid Cloud infrastructure," Journal of Parallel and Distributed  Computing, Vol. 72(10), pp.1318-1331, October 2012.

 \bibitem{chaisiri2012optimization}
S. Chaisiri, Bu-Sung Lee and D. Niyato, "Optimization of Resource Provisioning Cost in Cloud Computing," Services Computing, IEEE Transactions on , vol.5, no.2, pp.164,177, April-June 2012.


 \bibitem{lucas2012scheduling}
 J. Lucas-Simarro, R. Moreno-Vozmediano, R. Montero, and I. Llorente, "Scheduling strategies for optimal service deployment across multiple clouds," Future Generation Computer Systems, vol.29, no. 6, pp.1431-1441, August 2013. 

 \bibitem{breitgand2011policy}
 D. Breitgand, A. Marashini, and J. Tordsson, "Policy-driven service placement optimization in federated clouds,"' IBM Research Division, Tech. Rep, 2011.
 2011.

 \bibitem{vecchiola2012deadline}
C. Vecchiola, R. Calheiros, D. Karunamoorthy, and R. Buyya, "Deadline-driven provisioning of resources for scientific applications in hybrid clouds with
 Aneka," Future Generation Computer Systems, vol. 28, no. 1, pp.58-65, January 2012.

 \bibitem{vecchiola2009aneka}
C. Vecchiola, X. Chu, and R. Buyya, "Aneka: a software platform for .NET-based cloud computing," High Speed and Large Scale Scientific Computing, pp.267--295, 2009.

 \bibitem{wright2012constraints}
P. Wright, Y. L. Sun, T. Harmer, A. Keenan, A. Stewart, A., and  R. Perrott, "A constraints-based resource discovery model for multi-provider cloud environments," Journal of Cloud Computing: Advances, Systems and Applications, vol. 1, no. 1, 2012.

 \bibitem{calheiros2012coordinator}
 R. Calheiros, A. Toosi, C. Vecchiola, and R. Buyya "A coordinator for scaling elastic applications across multiple clouds," Future Generation Computer Systems Vol.28(8), pp.1350-1362, 2012.


 \end{thebibliography}
%
% <OR> manually copy in the resultant .bbl file
% set second argument of \begin to the number of references
% (used to reserve space for the reference number labels box)

% that's all folks
\end{document}